\documentclass{aa}
\usepackage{graphicx}

\def\lum{erg s$^{-1}$}

\begin{document}

\title{BL Lac Identification for the Ultraluminous X-ray Source Observed in
the Direction of NGC~4698}

\author{L. Foschini\inst{1}, L.~C. Ho\inst{2}, N. Masetti\inst{1}, M.
Cappi\inst{1}, M. Dadina\inst{1}, L. Bassani\inst{1}, G. Malaguti\inst{1}, E.
Palazzi\inst{1},\\ G. Di Cocco\inst{1}, P. Martini\inst{2}, S.
Ravindranath\inst{2,3}, J.~B. Stephen\inst{1}, M. Trifoglio\inst{1}, and
F. Gianotti\inst{1}}

\institute{Istituto di Astrofisica Spaziale e Fisica Cosmica (IASF--CNR) --
Sezione di Bologna\thanks{Formerly Istituto TeSRE -- CNR.}, Via Gobetti 101,
I--40129, Bologna (Italy)
\and
The Observatories of the Carnegie Institution of Washington, 813 Santa Barbara
Street, Pasadena, CA 91101 (USA)
\and
Department of Astronomy, University of California, Berkeley, CA 94720, USA
}

\offprints{L. Foschini, email: \texttt{foschini@bo.iasf.cnr.it}.}
\date{Received 5 August 2002; Accepted 20 September 2002}

\abstract{We report the identification of the optical and radio counterparts
of the ultraluminous X-ray (ULX) source XMMU~J124825.9+083020 (NGC4698-ULX1).
The optical spectrum taken with the VLT yields a redshift of
$z=0.43$, which implies that the ULX is not associated with the nearby galaxy
NGC~4698.  The spectral energy distribution calculated from the available data
indicates that the source is likely to be a BL Lac object.  The possible
synchrotron peak at X-ray energies suggests that this source may be a
$\gamma$-ray emitter.
\keywords{galaxies: active --- galaxies: Seyfert --- galaxies: BL Lacertae
objects: general}}

\titlerunning{ULX Identification in NGC 4698}
\authorrunning{L. Foschini et al.}

\maketitle

\section{Introduction}

In recent years, the improved imaging capabilities and increased sensitivity
of \emph{ROSAT}, \emph{Chandra} and \emph{XMM-Newton} have allowed us to
effectively study discrete sources in nearby galaxies beyond the Local Group.
Particularly intriguing is the discovery of off-nuclear X-ray sources with
luminosities well above the Eddington limit for a typical neutron star,
$\sim 10^{38}$ \lum, and up to $2\times 10^{40}$ \lum\ (e.g., Read et al.
1997; Colbert \& Mushotzky 1999; Roberts \& Warwick 2000; Makishima et al.
2000; Fabbiano et al. 2001).  These sources are typically called ultraluminous
X-ray sources (ULXs).  Despite much effort, little is presently known about
these sources. The identification of their optical counterparts is often
problematical when the sources are superposed against regions of high surface
brightness in the host galaxy.  To date, only two ULXs appear to have a clear
optical identification (Roberts et al. 2001; Wu et al. 2002), while for others
it has been possible to study only the nearby environment (Pakull \& Mirioni
2002; Wang 2002).

We have started a search for ULXs in a sample of nearby galaxies (Foschini et
al.  2002).  This paper concerns follow-up observations of the ULX in
NGC~4698.  We discuss the radio and optical counterparts and give a redshift
estimate.  We identify the source with a background source, most likely a
BL Lac object at $z = 0.43$.

\begin{figure*}[ht]
\begin{center}
\includegraphics[width=13cm]{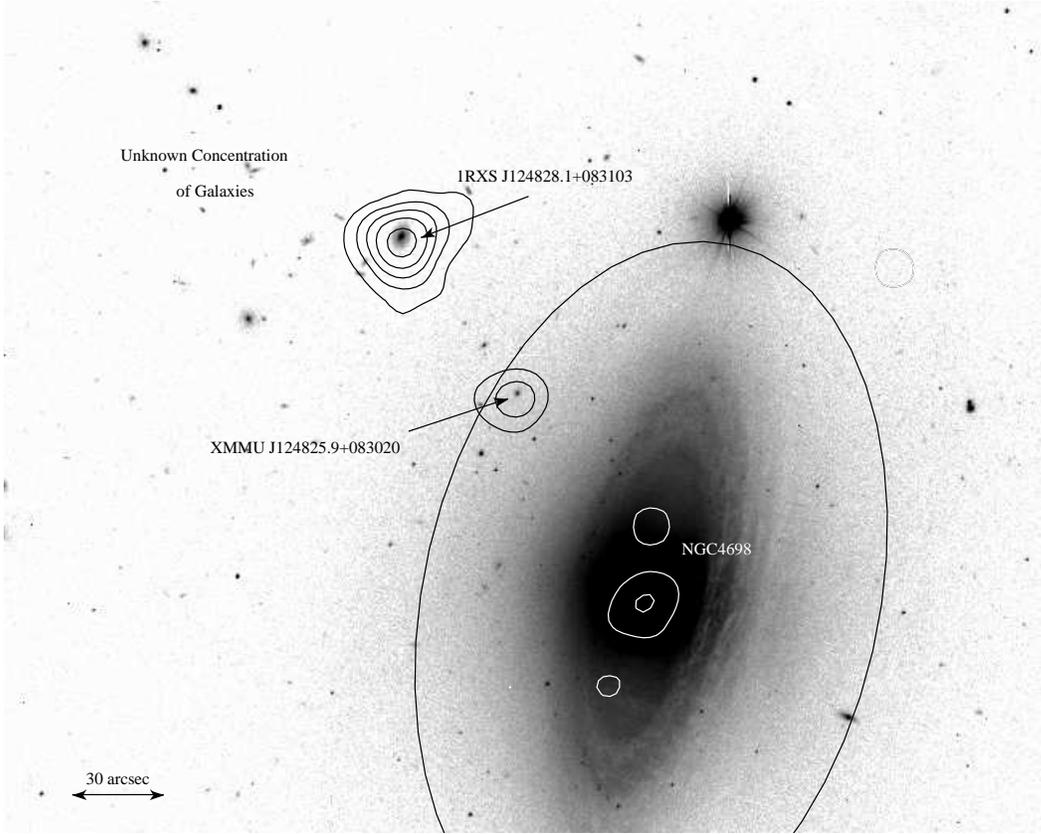}
\end{center}
\caption{$R$-band image from VLT-U3/FORS1, superimposed in contours the
smoothed image from the \emph{XMM-Newton} EPIC-MOS2 data in the $0.5-10$~keV
energy band (white over NGC~4698, black in the remaining field). North is up
and East to the left. The $D_{25}$ ellipse (3\farcm8) is shown for comparison.}
\label{fig1}
\end{figure*}

\section{X-ray data (\emph{XMM-Newton})}
NGC~4698 is an Sab spiral galaxy located in the Virgo cluster
($d=16.8$~Mpc).  It hosts an active nucleus, classified as a Seyfert 1.9 by
Ho et al. (1997).  The galaxy was observed on 16 December 2001 using the
European Photon Imaging Camera (EPIC) on board the \emph{XMM-Newton}
satellite. EPIC is composed of two instruments: the PN-CCD camera (Str\"uder
et al. 2001) and two MOS-CCD detectors (Turner et al. 2001).  The effective
exposure time was $9.2$~ks.

\begin{figure}[ht]
\begin{center}
\includegraphics[width=6.5cm]{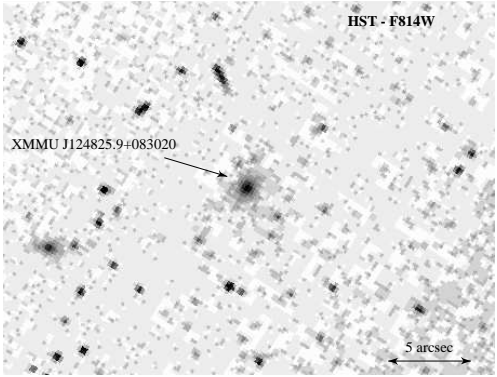}
\end{center}
\caption{Archival {\it Hubble Space Telescope}\ image of the region around
XMMU~J$124825.9+083020$ taken in the F814W filter. North is up and East to the
left.}
\label{fig2}
\end{figure}

\begin{table}[ht]
\caption{Results from the fit of the X-ray data. Columns: (1) Model: power law
(PL), black body (BB), bremsstrahlung (BR), multicolor black body disk (MCD);
(2) free parameter of the model: photon index $\Gamma$ for PL, temperature
(keV) for BB and BR, temperature (keV) at the inner disk for MCD; (3) $\chi^2$
and degrees of freedom of the spectral fitting; (4) flux in the $0.5-10$ keV
band ($10^{-14}$~erg cm$^{-2}$ s$^{-1}$); (5) X-ray luminosity in the $0.5-10$
keV band ($10^{43}$ \lum) calculated for $z=0.43$,
$H_{0}=75$~km~s$^{-1}$~Mpc$^{-1}$ and $q_{0}=0.5$.  No absorption is required
in excess of the the Galactic column, $N_{\rm H} =1.87\times 10^{20}$
cm$^{-2}$.  The uncertainties in the parameters are at the 90\% confidence
level.}
\centering
\begin{tabular}{lcccc}
\hline
Model & Parameter & $\chi^2$/d.o.f. & $F_{X}$ & $L_{X}$\\
(1)   & (2)       & (3)       & (4)             & (5) \\
\hline
PL    & $\Gamma=2.0\pm 0.2$ & 18.3/24    & 8.6  & 3.9\\
BB    & $kT=0.35\pm 0.05$   & 34.5/24    & 4.5  & {}\\
BR    &  $kT=2\pm 1$        & 19.0/24    & 6.7  & {}\\
MCD   & $kT_{in}= 0.6\pm 0.1$& 23.7/24   & 5.4  & {}\\
\hline
\end{tabular}
\label{tab:xdata}
\end{table}

We find only one ULX apparently associated with NGC~4698 (Foschini et al.
2002), located $73\arcsec$ from the optical nucleus of the galaxy, at $\alpha =
12^h48^m25.9^s$ and $\delta = +08^{\circ}30\arcmin20\arcsec$ (J2000). The
point source centroid is measured to precision of better than $0\farcs1$,
and the absolute pointing uncertainty is $< 4\arcsec$ (Jansen et al. 2001). We
detected $198$ photons with MOS1, in a circle of radius $30\arcsec$, $249$
with MOS2, and $414$ with PN.  These statistics are sufficient to perform
spectral fitting with simple models.  The ones we use are the power law (PL),
black body (BB), bremsstrahlung (BR), and the multicolor black body disk (MCD,
\emph{diskbb} in Xspec) by Mitsuda et al.  (1984).  Fluxes were corrected
according to the energy encircled fraction (Ghizzardi 2001).

For the processing, screening, and analysis of the data we used the standard
tools in the XMM-SAS software (v. 5.2) and HEAsoft Xspec (v. 11.0.1).
Correction for vignetting has not been applied because the source is close to
the center of the field of view ($< 2\arcmin$) and most of the detected
photons have energies less than $5$~keV (see Lumb 2002).

\begin{figure}[ht]
\begin{center}
\includegraphics[angle=270, width=8.0cm]{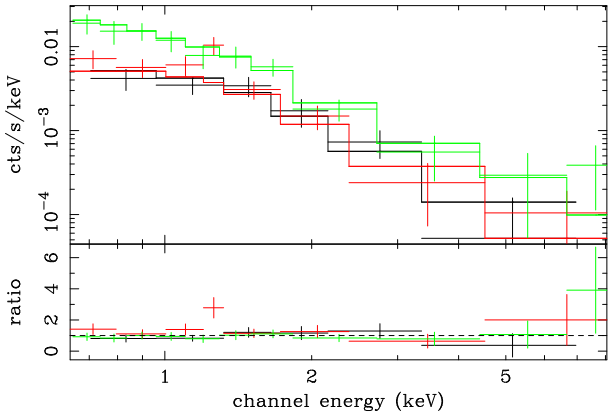}
\end{center}
\caption{\emph{XMM-Newton} spectrum of XMMU~J$124825.9+083020$. Data are from
MOS1, MOS2, and PN.}
\label{figxmmspe}
\end{figure}

The best fit (Fig.~\ref{figxmmspe}) is found using the power-law model with
$\Gamma = 2.0\pm 0.2$ ($\chi^2=18.3$, $\nu=24$), giving a flux of
$8.6\times 10^{-14}$~erg cm$^{-2}$ s$^{-1}$.  If the source is located in
NGC~4698, for which we assume a distance of 16.8 Mpc, the corresponding
luminosity, assuming isotropic emission, is $\sim 3\times 10^{39}$ \lum.  For
$z = 0.43$, as discussed below, the luminosity becomes $3.9\times 10^{43}$
\lum.  Additional information on the fits using the other models is given in
Table~\ref{tab:xdata}.

\section{Radio data (VLA)}

The VLA observations of NGC~4698 were performed by Ho \& Ulvestad (2001) on
29 August 1999 (20~cm) and 31 October 1999 (6~cm). The source detection at
6~cm was reported by Ho \& Ulvestad (2001); the flux density is $1.13$~mJy and
the background noise is $0.072$~mJy beam$^{-1}$.  The source, which appears
largely unresolved at a resolution of $\sim 1\arcsec$, is located at
$\alpha = 12^h48^m25.9^s$ and $\delta = +08^{\circ}30\arcmin21\arcsec$
(J2000); the radio position is accurate to $\sim 0\farcs1$.

A reanalysis of the VLA data detected the source also at 20~cm.  It is
unresolved with the 1$\arcsec$ beam, and its 20~cm position is consistent with
that measured at 6~cm.  We find a flux density of $0.34$~mJy, with a
background noise of $0.032$~mJy beam$^{-1}$.  The spectral index, defined as
$S_{\nu}\propto \nu^\alpha$, is $\alpha_{r}\approx 1.0$.

\section{Optical data}
\subsection{Digitized Sky Survey (DSS)}
The optical counterpart of the source is visible on the DSS and the {\it HST}\
observations of NGC~4698.  We are certain of the identification from the
relative offset between the nucleus and the optical counterpart.  From the
Automatic Plate Measurement
(APM\footnote{\texttt{http://www.ast.cam.ac.uk/$\sim$apmcat/}.})
catalog, the point source has $R = 18.9$ mag and $B = 21.9$ mag, indicating
a very red color ($B-R=3.0$ mag). The original observation was made with the
48~inch Schmidt telescope at Palomar Observatory on 17 February 1950.

\subsection{Hubble Space Telescope (HST)}
We retrieved the calibrated F450W, F606W, and F814W WFPC2 images of NGC~4698
from the HST Archive\footnote{Based on observations made with the NASA/ESA
Hubble Space Telescope, obtained from the data archive at the Space Telescope
Institute.  STScI is operated by the association of Universities for Research
in Astronomy, Inc. under the NASA contract  NAS 5-26555.} to perform photometry
of the optical counterpart of the source. The observations with the F450W and
F814W filters were performed on 10 August 2001, while the observation with
F606W was taken on 31 March 2000. In all the filters, the source is located
on the WF detector, which has a scale of 0\farcs1 per pixel. The total exposure
time is 460~s ($2\times 230$~s) in the F450W and F814W filters and 600~s
(400~s and 200~s) in the F606W filter. We combined the separate images taken in
each filter using the {\it crrej}\ task in IRAF to reject cosmic rays.
Aperture photometry was done using the tasks available in the {\it apphot}\
package.  The integrated magnitudes, determined using the aperture growth
curve method, were transformed to the Vega magnitude system using the
zeropoints given in the {\it HST}\ Data Handbook. We find $m_{\rm F450W} =
21.1\pm 0.1$ mag, $m_{\rm F606W} = 19.9\pm 0.1$ mag, and $m_{\rm F814W} =
18.6\pm 0.1$ mag. In the Johnson-Kron-Cousins system these values translate
into $B=21.3\pm 0.1$ mag, $R=19.6\pm 0.1$ mag, and $I=18.5\pm 0.1$ mag.

\subsection{Las Campanas (Magellan)}
XMMU~J$124825.9+083020$ was observed for 900~s ($3\times 300$~s) on 6 April
2002 with the LDSS-2 spectrograph (Allington--Smith et al. 1994) on the 6.5~m
Baade telescope at Las Campanas Observatory. We observed through a $1\farcs25$
longslit in $0\farcs9$ seeing, with the slit oriented at the parallactic
angle. We used the medium red grism, which gives a dispersion of $5.3$~\AA\
pixel$^{-1}$. The FWHM spectral resolution was $16$~\AA, and the spatial
resolution was $0\farcs38$ pixel$^{-1}$.

Basic data reduction was performed using the IRAF package. The individual
spectroscopic frames were corrected for overscan, flat-fielded using domeflats
and summed to obtain the final object frame. The spectral extraction was done
by summing the counts within an aperture of $6$ pixels ($2\farcs3$).
Wavelength calibration was achieved using a polynomial fit to the lines in the
sky spectrum.

The spectrum in the range $5000-7000$~\AA\ is approximately described by the
form $f_{\lambda}\approx \lambda^{4}$.

\begin{figure*}[ht]
\begin{center}
\includegraphics[angle=270,width=18cm]{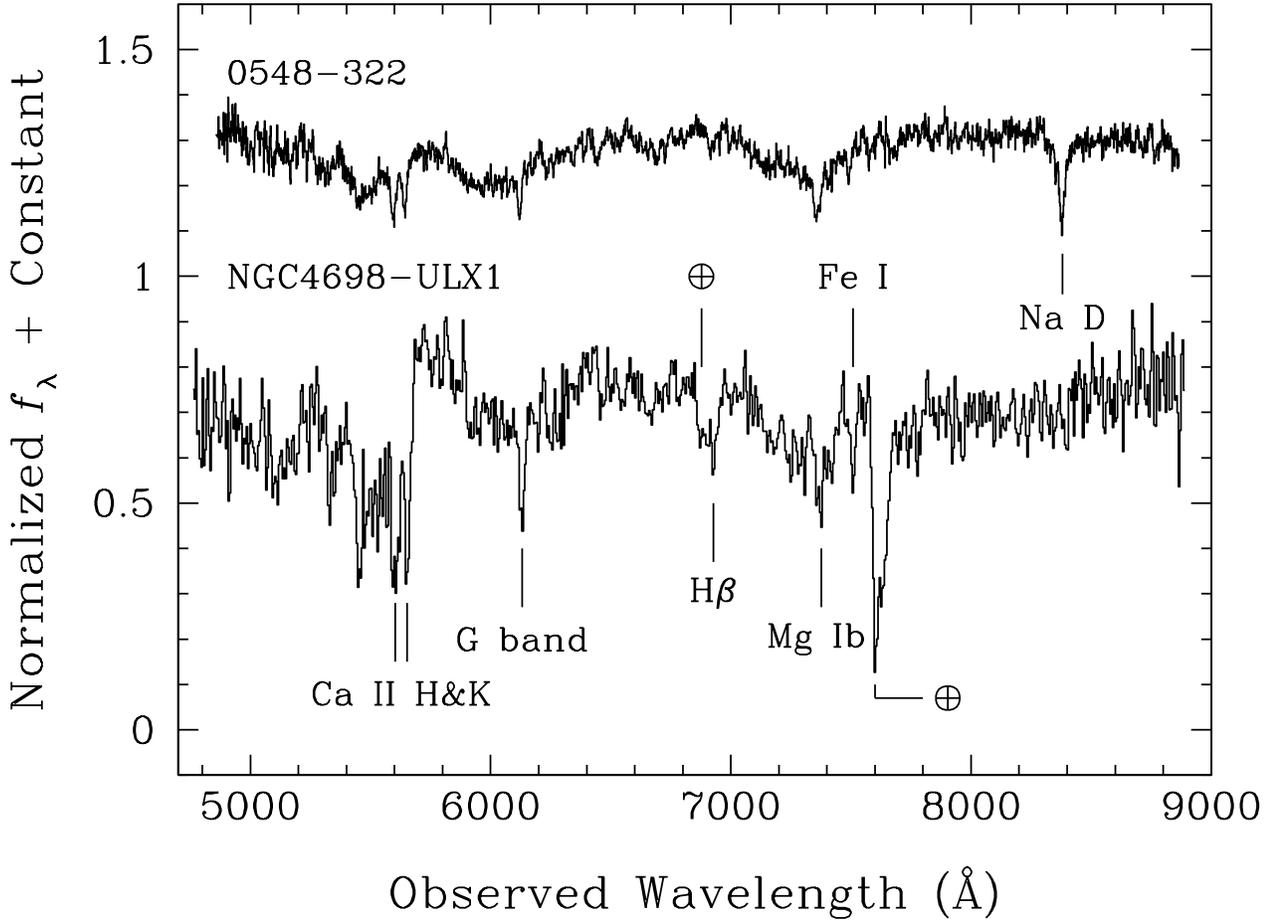}
\end{center}
\caption{Normalized VLT spectrum with identifications of absorption lines
marked.  Telluric absorption lines are marked with the symbol $\oplus$.  Shown
for comparison is the spectrum of the BL Lac object $0548-322$ (Barth et al.
2002), redshifted to $z=0.43$.}
\label{fig3}
\end{figure*}

\subsection{Very Large Telescope (VLT)}
VLT observations\footnote{Based on observations made with ESO Telescopes at
the Paranal Observatory under programme ID~$269.D-5014$.} were performed with
VLT-UT3 (Melipal) plus the FORS1 spectrograph.  FORS1 is equipped with a
$2048\times2048$ pixel Tektronix CCD, which covers a $6\farcm8\times 6\farcm8$
field in the standard resolution imaging mode with a scale of $0\farcs 2$
pixel$^{-1}$.

Two spectra, both with an exposure time of 570~s, were acquired on 21 April
2002, starting at 4:52:50 UT.  The spectra were acquired using Grism \#150I
plus order separator GG435, which avoids overlapping of spectral orders over
a given wavelength; this limited the spectral range to $4500-9000$~\AA.  The
slit width was $1\arcsec$ for both spectra, and this setup secured a final
dispersion of $5.5$~\AA~pixel$^{-1}$, corresponding to a FWHM resolution of
13 \AA.  Before the spectroscopic observation, a one minute $R$-band acquisition
image was obtained on the same night starting at 04:42:52 UT under very good
seeing conditions ($\sim$0$\farcs$6). The object
appeared clearly elliptical and extended, with a bright core and a fuzzy
halo.  Moreover, an irregular spot located around 1$\arcsec$ north of the
object core is apparent in the VLT image.

After correction for flat-field and bias, the spectra were background
subtracted and optimally extracted (Horne 1986) using IRAF.  He-Ne-Ar and
Hg-Cd lamps were used for wavelength calibration.  The wavelength calibration
was checked against the position of night sky lines; the typical error was
0.5 \AA.  Finally, the two spectra were stacked together in order to increase
the signal-to-noise ratio.  We encountered problems during observation of
the spectrophotometric standard, and hence our spectra are not flux calibrated.

Fig.~\ref{fig3} shows the normalized VLT spectrum, with several absorption
lines identified [see Laurent-Muehleisen et al. (1998) for a discussion on
optical identification of BL Lacs].  The detected lines include Ca~II H\&K
$\lambda\lambda 3933, 3968$, the G band $\lambda 4304$, H$\beta$
$\lambda 4861$, and Fe~I $\lambda 5270$: they all indicate $z \approx 0.43$.
The line strengths are heavily diluted by the featureless continuum, but the
spectrum shares close similarity to that of the BL Lac object $0548-322$
(Barth et al. 2002).  We shifted the spectrum of $0548-322$, which has a
redshift of 0.069, to $z = 0.43$.

\begin{figure}[ht]
\begin{center}
\includegraphics[angle=270, width=8.0cm]{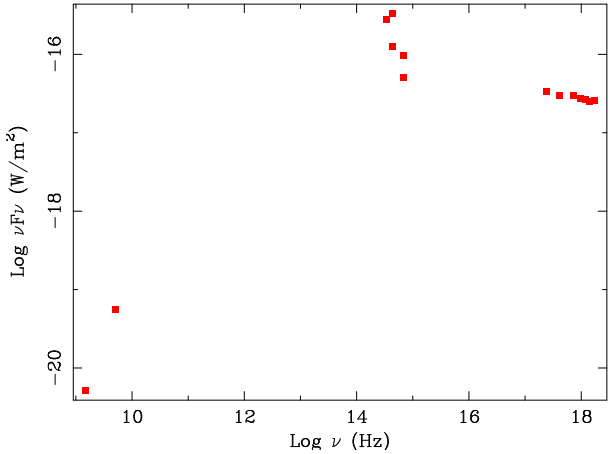}
\end{center}
\caption{The spectral energy distribution of XMMU~J124825.9+083020, assembled
from data taken with the VLA, DSS, {\it HST}, and {\it XMM}.}
\label{fig4}
\end{figure}

\section{Discussion}
\subsection{Classification}
The absorption features detected in the VLT spectrum clearly places
XMMU~J124825.9+083020 at $z=0.43$.  We identify  the source as a
BL Lac object, for the following reasons.  (1) We detect no
emission lines; the upper limit to any emission feature is $\sim$5 \AA.
(2) The stellar features superposed on the featureless continuum
have strengths generally consistent with those of other BL Lac objects.  We
demonstrate this concretely by comparing a redshifted spectrum of the
BL Lac object $0548-322$ (Fig.~\ref{fig3}).  The break contrast at 4000 \AA\
is $\sim$29\%, again similar to other BL Lac objects (Laurent-Muehleisen
et al. 1998).  (3) The optical continuum slope measured in the
Magellan spectrum is consistent with those of high-frequency peaked
BL Lac objects (e.g., Stickel et al. 1993).  And (4) the multiwavelength
spectral energy distribution (Fig.~\ref{fig4}), although not assembled from
simultaneous observations, is highly reminiscent of those of BL Lac objects.

By using two-point spectral indices, namely the radio-to-optical
$\alpha_{\mathrm{ro}}$ and optical-to-X-ray $\alpha_{\mathrm{ox}}$, it is
possible to show that different objects populate different regions of the
$\alpha_{\mathrm{ro}} - \alpha_{\mathrm{ox}}$ plane (e.g., Brinkmann et al.
1997; Laurent-Muehleisen et al. 1999). Another method has been suggested by
Maccacaro et al. (1988), who proposed a nomograph to link the X-ray flux in
the energy band $0.3-3.5$~keV and the visual magnitude.  The values of
$\alpha_{\mathrm{ro}}$ and $\alpha_{\mathrm{ox}}$ for the present source are
$0.42$ and $0.95$, respectively, thus placing it in the region of X-ray
selected BL Lacs (Brinkmann et al. 1997), or high-energy peaked BL Lacs in the
$\alpha_{\mathrm{ro}}-\alpha_{\mathrm{ox}}$ diagram of Laurent-Muehleisen
et al.~(1999). The nomograph of Maccacaro et al. (1988) gives a ratio
$f_{\mathrm{x}}/f_{\mathrm{v}}$ between $1.3$ and $3.8$ (depending on whether
we use the {\it HST}\ optical magnitudes through the F606W or F450W filter,
respectively), in the regime of AGNs and BL Lacs.

The spectral indices can be used to deduce some general properties of the
dominant radiation mechanism.  If $\alpha_{\mathrm{x}} \leq
\alpha_{\mathrm{r}}$, the source may exhibit relativistic beaming, while if
$\alpha_{\mathrm{x}} > \alpha_{\mathrm{r}}$, it may fulfill the conditions of
the homogeneous synchrotron model (Harris \& Krawczynski 2002). In our case,
we have $\alpha_{\mathrm{x}} \approx \alpha_{\mathrm{r}} \approx 1$, so that
we cannot clearly discriminate between these two cases.  If the source
is a high-frequency peaked BL Lac, however, it is likely that the homogeneous
synchrotron model is more applicable.
This is confirmed by the $\alpha_{\mathrm{xox}}$ test of Sambruna et al.
(1996): in this case, the difference by
$\alpha_{\mathrm{ox}}-\alpha_{\mathrm{x}}=\alpha_{\mathrm{xox}}$ is
approximately equal to zero, so avoiding a clear discrimination between the
physical mechanism of the source.

The spectral energy distribution (Fig.~\ref{fig4}) is comparable to those of
BL Lac objects peaked in the $\gamma$-ray domain (e.g., Fossati et al. 1998),
but the third EGRET catalog (Hartman et al. 1999) does not have any source
within several degrees of XMMU~J$124825.9+083020$.

\subsection{Search for ULXs and Contamination with Background Objects}
This research is a useful demonstration of just how difficult it is to
identify the physical nature of ULXs.  It shows the vital importance of
redshift determinations.  Although we had a lot of photometric data on
XMMU~J$124825.9+083020$, by themselves they were insufficient to clearly
establish whether the source belongs to NGC~4698 or is a background object.
In the first case, the source would have been something similar to
a microquasar (e.g., Mirabel \& Rodr\'{\i}guez 1999), probably located in a
globular cluster. Indeed, the optical image of the source appeared to be
slightly extended (angular extent $\sim 3\arcsec$), and its position with
respect to NGC~4698 suggested that it could be a globular cluster, albeit an
unusually large one. The possibility of an accreting black hole in a
globular cluster is not so remote: \emph{Chandra} observations of the globular
cluster system of NGC~4472 show that about $40$\% of the bright low mass X-ray
binaries are associated with optically identified globular clusters (Kundu et
al. 2002). In addition, the X-ray luminosity function shows a break near
$3\times 10^{38}$ \lum, suggesting that the brightest X-ray binaries are
accreting black holes (Kundu et al. 2002), perhaps microquasars.

\emph{XMM-Newton} observations of the Lockman Hole (Hasinger et al. 2001)
show that the number of background sources in the energy band $0.5-2$~keV with
flux greater than $4.0\times 10^{-14}$~erg cm$^{-2}$ s$^{-1}$ (the best fit
value from Tab.~\ref{tab:xdata}) is $15$ deg$^{-2}$.  In the energy band
$2-10$~keV, there are 40 sources deg$^{-2}$ with flux higher than
$4.6\times 10^{-14}$~erg cm$^{-2}$ s$^{-1}$.  Assuming the same $\log N -
\log S$ relation, and considering that the $D_{25}$ area of NGC~4698 is about
$7.9$~arcmin$^2$, we expect $0.08$ background objects in the $2-10$~keV energy
band and $0.03$ in the energy band $0.5-2$~keV. However, despite these low
values, we have found, in the present case, that the only ULX is
a background AGN.

The above calculations could be underestimated in the present case because
the VLT images show an unknown concentration of galaxies north-east of
NGC~4698 (see Fig.~\ref{fig1}), thus suggesting the possibility of a
statistically meaningful excess of background sources. However, no additional
X-ray sources is seen in the present \emph{XMM-Newton} observation. Perhaps, a
longer exposure may reveal soft X-ray emission or additional sources, if
NGC 4698 lies along the line of sight to a galaxy cluster.  The three X-ray
sources identified to date, however, have three redshifts:
XMMU~J$124825.9+083020$ has $z=0.43$, NGC~4698 has $z=0.0033$, and the
{\it ROSAT}\ source 1RXS~J$124828.1+083103$ has been recently identified with
a Seyfert nucleus at $z=0.12$ (Xu et al. 2001). Therefore, these three sources
are not members of a single cluster.

\section{Final Remarks}
We have presented a multiwavelength analysis of the ultraluminous X-ray source
XMMU~J124825.9+083020 observed in the direction of NGC~4698.  The availability
of X-ray, optical, and radio data makes it possible to construct the spectral
energy distribution, but the crucial information to decipher the true
nature of the source was provided by the optical spectrum.  We show that
the absorption features are unambiguously associated with a background
source at $z=0.43$.  We argue that the source is most likely a BL Lac object,
most likely of the high-frequency peaked category, based on the absence of
emission lines, the dilution of the absorption features, the slope of the
optical continuum, and the overall shape of the spectral energy distribution.

\begin{acknowledgements}
We would like to thank A. Pizzella for useful discussion about NGC~4698, the
Service Mode Observations personnel at ESO, and in particular M. Romaniello,
for the help in the preparation of VLT observations.  This work is based on
observations obtained with \emph{XMM-Newton}, an ESA science mission with
instruments and contributions directly funded by ESA Member States and the USA
(NASA).  This research has made use of the NASA Astrophysics Data System
Abstract Service and of the NASA/IPAC Extragalactic Database (NED), which is
operated by the Jet Propulsion Laboratory, California Institute of Technology,
under contract with the National Aeronautics and Space Administration.
We acknowledge the partial support of the Italian Space Agency (ASI) to this
research.  L.~C.~H. is grateful for financial support from the Carnegie
Institution of Washington and from NASA grants.
\end{acknowledgements}

\end{document}